\documentclass[twocolumn, amsmath,amssymb,aps]{revtex4-1}

\usepackage{verbatim}
\usepackage{graphicx}
\usepackage{bm} % bold math 
\usepackage{ifpdf} % This lets us get figures right both in pdf and ps versions
\usepackage{dcolumn}  % Align table columns on decimal point
\usepackage{epsfig}
\usepackage{color}
\usepackage[usenames,dvipsnames]{xcolor}
\usepackage{siunitx}

\newcommand{\trm}[1]{{\textrm{#1}}}
\def\rmd{{\mathrm{d}}}

\def\rme{{\mathrm{e}}}
\def\lB{\ell_{\mathrm{B}}}
\newcommand{\Av}[1]{{\bf #1}}
\newcommand{\kB}{k_\mathrm{B}}
\def\Eq{Eq.}

\def\Fig{Fig.}

\newcommand{\cF}{{\cal F}}
\newcommand{\cZ}{{\cal Z}}
\newcommand{\cB}{{\cal B}}

\newcommand{\celsius}{\ensuremath{^\circ}C}

\begin{document}

\title{Interactions between Charged Particles with bathing Multivalent Counterions: Experiments {\it vs} Dressed Ion Theory}
\author{Matej Kandu\v c$^1$}
\author{Mohsen Moazzami-Gudarzi$^2$}
\author{Valentina Valmacco$^2$}
\author{Rudolf Podgornik$^{3,4}$}
\email{E-mail: \texttt{rudolf.podgornik@ijs.si}}
\author{Gregor Trefalt$^2$}
\email{E-mail: \texttt{gregor.trefalt@unige.ch}}

\affiliation{$^1$Soft Matter and Functional Materials, Helmholtz-Zentrum Berlin f\"ur Materialien und Energie, Hahn-Meitner-Platz 1, D-14109 Berlin, Germany}
\affiliation{$^2$Department of Inorganic and Analytical Chemistry, University of Geneva, Sciences II, 30 Quai Ernest-Ansermet, 1205 Geneva,
Switzerland}
\affiliation{$^3$Department of Theoretical Physics, J. Stefan Institute, 1000 Ljubljana, Slovenia}
\affiliation{$^4$Department of Physics, Faculty of Mathematics and Physics, University of Ljubljana, 1000 Ljubljana, Slovenia}

% \date{\today}

\begin{abstract}
We compare the recent experimentally measured forces between charged colloidal particles, as well as their effective surface potentials (surface charge) in the presence of multivalent counterions in a bathing monovalent salt solution, with the predictions of the dressed ion theory of strongly charged colloidal systems. The benchmark for comparison is provided by the DLVO theory and the deviations from its predictions at small separations are taken as an indication of the additional non-DLVO attractions that can be fitted by an additional phenomenological exponential term. The parameters characterizing this non-DLVO exponential term as well as the dependencies of the effective potential on the counterion concentration and valency predicted by the dressed ion theory are well within the experimental values. This suggests that the deviations from the DLVO theory are probably caused by ion correlations as formalized within the dressed ion theory. 
\end{abstract}

\maketitle

%%%%%%%%%%%%%%%%%%%%%
%%% Introduction %%%%
%%%%%%%%%%%%%%%%%%%%%

\section{Introduction}

Electrolyte solutions bathing charged colloids, macromolecules, and macroions that are composed of asymmetric ionic mixtures with mono- and multivalent ions, can drastically alter the long-range interactions between them and consequently modify their solution behavior. Often these modifications brought by asymmetric ionic mixtures challenge our fundamental understanding of these systems standardly based on the Derjaguin-Landau-Verwey-Overbeek (DLVO) paradigm~\cite{Andelman}, and make it difficult to predict and control their behavior~\cite{Perspective}. A diverse range of phenomena display a crucial dependence on the detailed identity of the bathing solution in terms of its ionic composition. Such phenomena include a plethora of biological systems ranging from the formation of large aggregates of like-charged biopolymers, such as microtubules, F-actin, and DNA in the bulk electrolyte solutions~\cite{Wong:2010, Nguyen}, and then all the way to the packaging of DNA inside viral shells~\cite{Podgornik2016,Nguyen1} and in the eukaryotic chromatin\cite{Schiessel2003}. Electrostaic interactions are also important in many technological processes such as waste water treatment, paper making, and concrete hardening~\cite{Holm2012, Bolto2007, Iselau2015, Labbez2006}.

The classical understanding of electrostatic interactions mediated by electrolyte solutions is embodied in the DLVO theory~\cite{Derjaguin1941,Verwey1948,Andelman,Monica1}, which decomposes the total interactions between charged macroions in the presence of mobile solution ions into two principal components, viz. the van der Waals (vdW) forces and the double layer (DL) forces. The former consistently provide attractive long-range interactions between dielectrically similar bodies~\cite{Parsegianbook,RMP2016}, while the latter are typically
framed within the mean-field Poisson–Boltzmann (PB) theory~\cite{Andelman}. It is in fact the mean-field nature of the PB description that fails to address the ion–ion correlations in the electrolyte solutions and thereby also fails to properly describe the electrostatic interactions in situations where the electrostatic coupling is strong. This is certainly the case with large macroion surface charges and/or in the presence of multivalent solution ions~\cite{Kanduc2010, Grosberg, Levin, Boroudjerdi}. The theoretical aspects of the ion correlation effects have been addressed exactly with explicitly solvable low-dimensional models~\cite{Dean1998,Dean2009, Demery2009} and with simulations in the framework of different (coarse grained) electrolyte solution models~\cite{SIMUL, Jho2008, Kanduc2008,Nguyen}. Approximate approaches at various levels of description~\cite{Kanduc2010} were also used: based on the integral-equation methods~\cite{Kjellander1, Kjellander2}, perturbative improvement of the mean-field theory including loop expansions and other Gaussian- fluctuations approximations\cite{PodgornikZeks,Joanny,Pincus,Ha-Liu,Lukatsky1,Lukatsky2, Golestanian,Lau2002}, variational methods~\cite{Orland,Buyukdagli1,Buyukdagli2,Wang1,Wang2}, local density functional theory~\cite{Robbins,Diehl,Forsman}, the strong coupling expansion~\cite{Boroudjerdi,Naji2005}, dressed ion expansion~\cite{Kanduc2010,Kanduc2011}, and the Wigner crystal (low temperature) expansion~\cite{Grosberg,Samaj}. While there is obviously no shortage of theoretical approaches, less effort was dedicated to systematic investigations of the applicability of the various theories to different well-defined experimental situations characterized by complicated multicomponent bathing electrolyte solutions.

In fact, experimental investigation of the interactions between charged colloids, macromolecules and {macroions} or charged macromolecular surfaces across aqueous electrolyte solutions have reached a stage where they can be routinely and accurately measured easily down to sub-nanometer separations. This development was enabled with the advancement of experimental techniques such as osmotic-stress~(OS) methodology~\cite{Rau,yasar2014},  surface force apparatus~(SFA)~\cite{Israelachvili2010, Espinosa-Marzal2012}, colloidal probe atomic force microscopy~(AFM)~\cite{Ducker1991, Butt1991}, total internal reflection microscopy (TIRM)~\cite{Prieve1999, vonGrunberg2001}, and optical tweezers~\cite{Gutsche2007, Crocker1994}. Direct force measurement in the presence of multivalent ions received increased interest lately, specifically in connection with the assessment of the validity of the mean-field PB description and/or the DLVO paradigm in general~\cite{Perspective}. Investigations of the effect of multivalent ion solution date a while back, to the direct measurement of the interactions between two mica sheets in the presence of trivalent cations as studied in detail with the SFA methodology~\cite{Pashley1984}, as well as to the OS methodology used to deconvolute the interactions between highly charged biopolymers such as DNA in the presence of multivalent counterions, e.g.\ CoHex, spermine, and spermidine~\cite{Strey1,Rau,yasar2014}. More recently, colloidal probe AFM was set up to study the forces between a colloidal particle and a surface or two either similar or dissimilar colloidal particles  of negative or positive surface charge in the presence of multivalent counterions~\cite{Besteman2004, Zohar2006, Dishon2011, MoazzamiGudarzi2015, MontesRuiz-Cabello2013, MontesRuiz-Cabello2014b, Valmacco2016, Trefalt2017}. Some measurements were also done in the presence of multivalent coions, however these interactions are not the focus of the present paper~\cite{MontesRuiz-Cabello2015}. Sivan and coworkers~\cite{Zohar2006, Dishon2011} have observed an exponential short-range attraction between two silica surfaces in the presence of trivalent counterions near charge neutralization, with decay lengths of a few nanometers that were stronger than the estimated vdW forces. Further experimental work performed within the Borkovec group concentrated on a more detailed comparison with the mean-field PB calculations of the DL interactions~\cite{Sinha2013, MoazzamiGudarzi2015, MontesRuiz-Cabello2013, MontesRuiz-Cabello2014b, Valmacco2016}. The main conclusions stemming from these experimental endeavours elucidating the forces in the presence of multivalent counterions can be summarized as follows: The long-range separation behavior of the experimental forces can be accurately described by the PB theory, however, the effective or the renormalized surface potentials and/or charges must be used to accurately describe the data~\cite{Sinha2013, MoazzamiGudarzi2015, MontesRuiz-Cabello2013, MontesRuiz-Cabello2014b, Valmacco2016}. In many situations, these renormalized potentials are close to the surface potential values measured by electrophoretic mobility~\cite{Sinha2013, MoazzamiGudarzi2015, MontesRuiz-Cabello2013, MontesRuiz-Cabello2014b}. Moreover, multivalent ions often induce {\sl charge neutralization} or even {\sl charge reversal} as observed from both electrophoretic mobility and direct force measurements~\cite{Zohar2006, Sinha2013, MoazzamiGudarzi2015, MontesRuiz-Cabello2013, MontesRuiz-Cabello2014b, Trefalt2016}. This type of behavior cannot be explained by the mean-field approach. Another  interesting observation is that at smaller separations, e.g.\ below a few nanometers, {\sl additional attractive forces} are observed that cannot be captured within the DLVO framework. While these attractive forces can be approximated phenomenologically by an exponential dependence on the separation, their physical origin remains elusive and their source could be and indeed has been variously attributed to ion--ion correlations, finite ionic size, image-charge interactions, charge heterogeneities, or charge fluctuations~\cite{Besteman2004, Zohar2006, Dishon2011, MoazzamiGudarzi2015, MontesRuiz-Cabello2013, MontesRuiz-Cabello2014b, Valmacco2016, Moazzami-Gudarzi2016}.

While both experimental as well as theoretical studies of strongly charged electrostatic systems have been obviously well researched (see above), there are very few, if indeed any, detailed, quantitative comparisons between them, that would indicate the range of validity and/or inadequacy of different theoretical models. It seems that the weakest point of a systematic comparison is the generally more complicated and varied composition of experimental setups, which typically contain an {\sl electrolyte mixture} composed of monovalent and multivalent salts, making many of the existing theoretical approaches too simplistic to start with. To infuse additional realism into the model system in its turn stipulates a price in the sense that analytical limits of the e.g.\ strong coupling or Wigner crystal types, are more difficult to access. In this respect, the recently introduced {\sl dressed counterion} theory~\cite{Kanduc2010,Kanduc2011}, in this work referred more generally as {\sl dressed ion}~(DI) theory (to be differentiated from the dressed ion theory as introduced by Kjellander et al., see Ref. \cite{Kjellander2}), presents a viable compromise with a sufficiently sophisticated bathing electrolyte model as well as simple analytical limiting laws, not too far removed from the simplicity, if more constrained validity, of the PB results.

Mixtures of multivalent ions in a bathing solution of monovalent ions with variable concentration is a particularly relevant realization of an electrolyte solution mediating electrostatic interactions between colloids. Because of the asymmetric nature of the electrolyte mixture---multivalent ions with monovalent salt ions---different approaches need to be invoked to describe its different components, since no single approximation scheme would be able to address all the charged solution components on the same level: while the strong coupling framework could be adequate for the multivalent counterions, it is not applicable to the monovalent salt. The converse is true for the mean-field PB framework. Based on these observations, a {\sl dressed counterion} framework was set up that allows a selective use of different approximation schemes for different components of the asymmetric electrolyte solution~\cite{Kanduc2010, Kanduc2011}. In this context, the system is assumed to be composed only of strongly coupled counterions and fixed colloidal charges that interact via screened pair potential. The multivalent counterions are therefore ``dressed" into the screening atmosphere of the monovalent salt ions as well as the remaining monovalent component of the multivalent salt, while concurrently the explicit salt degrees of freedom are integrated out on the linearized PB (i.e., Debye--H\"uckel) level and do not appear anymore in the partition function. This substantially simplifies the theory and leads to relatively simple expression for the interaction pressure between planar macromolecular and/or colloidal surfaces, which can be directly compared and fitted to the experiments.

In this work, we thus attempt to confront the DI theory predictions with experimentally measured forces between charged colloidal particles in the presence of multivalent counterions in a bathing monovalent salt solution medium. We expand the DI theory to the second virial order, which allows us to assess the regime of validity of the theory and provides additional explanations of the measured trends.
The mean-field PB theory is used as a benchmark case and the deviations at small distances are studied to elucidate the source of the additional non-DLVO attractions. Furthermore, the evolution of the effective surface potentials is calculated and compared to the experimentally determined potentials.

\section{Methods}

\subsection{Model system}
\begin{figure}[t]
\small
\centering
\includegraphics[width=8.5cm]{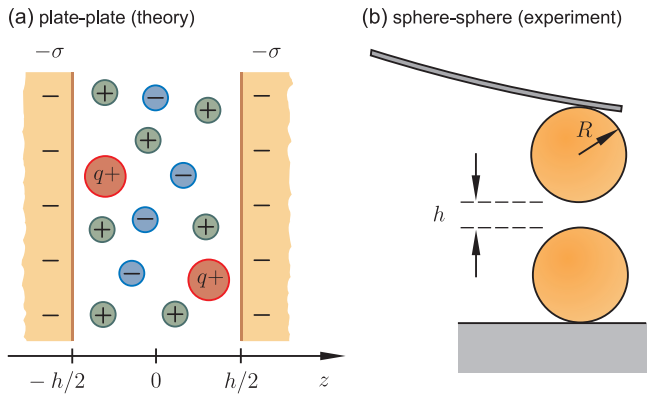}
\caption{Schematic illustration of (a)~plate--plate geometry of two charged surfaces with monovalent ions and multivalent countrions used in the theoretical calculations, and (b)~sphere--sphere geometry of two colloidal particles in the experimental colloidal probe measurements.
}
\label{fig:schematics}
\end{figure}

We compare experimental and theoretical results of interactions between charged surfaces. The interacting surfaces possess negative uniformly distributed surface charge $-\sigma$, bathing in an electrolyte solution containing symmetric 1:1 salt with bulk concentration  $n_0$, as well as an asymmetric multivalent salt $q$:1 of concentration $c_0$, where the multivalent ions of valency $q$ correspond to counterions with regard to the surfaces, see \Fig~\ref{fig:schematics}a. Thus the total bulk concentration of monovalent ($+1$) counterions is $n_0$, that of polyvalent ($+q$) counterions $c_0$, and monovalent ($-1$) coions $n_0+q c_0$.

While all the calculations were performed in the plate--plate geometry, which simplifies the theoretical formalism, all the experiments were performed in the sphere--sphere geometry, which simplifies the measurements (\Fig~\ref{fig:schematics}). The connection between the two, valid for closely apposed spheres, is provided through the {\sl Derjaguin approximation}~\cite{Derjaguin,Russel1989}
\begin{equation}
F(h) = 2\pi R_{\rm eff} f(h),
\label{eq:Derjaguin}
\end{equation}
where $F(h)$ is the force between two identical spheres of radius $R$, $f(h)$ the free energy per surface area between the plates, and $R_{\rm eff} = R/2$. The Derjaguin approximation is valid as long as the condition $h/R \ll 1$ is fulfilled, where $h$ is the surface--surface separation at the point of closest approach~\cite{Derjaguin}.

\subsection{Interaction decomposition}

Within the DLVO framework, the interactions are composed of electrostatic~(el) and vdW contributions~\cite{Russel1989}, so that the total interaction free energy per surface area $A$, $f = {\cal F}/A$, assumes the form
\begin{equation}
f(h) = f_{\rm el}(h) + f_{\rm vdW}(h).
\label{eq:force}
\end{equation}
where $h$ is the intersurface separation. At this point, it is important to recognize that for weakly charged systems the electrostatic component of the free energy, $f_{\rm el}$, is given by the PB approximation, while the vdW component acquires a screened zero Matsubara frequency term instead of the full Lifshitz expression ~\cite{Parsegianbook, RMP2016}. For strongly charged systems described by the dressed ion theory, the electrostatic component is given by the DI expression (see below) and the vdW component does not contain the zero Matsubara frequency term.

The higher-order terms in the Matsubara summation of the vdW interaction free energy are approximated by the standard Lifshitz non-retarded expression~\cite{Parsegianbook}
\begin{equation}
f_{\rm vdW} =  -\frac{H}{12\pi}\cdot\frac{1}{h^2}.
\label{eq:vdW}
\end{equation}
This form is  appropriate for separations not exceeding a few 10~nm. 
Note that the screening of the zero Matsubara term has only a limited effect on the effective Hamaker constant. The difference between the effective Hamaker constant for polystyrene particles across water at the low salt (1~mM) and high salt (1~M) is less than 20~\% at relevant separation distances~\cite{Elzbieciak-Wodka2014, Russel1989}.

\subsubsection{Poisson--Boltzmann theory}
In our first theoretical, DLVO approach, used as a benchmark, we base the electrostatic part of the interaction in \Eq~(\ref{eq:force}) on the mean-field PB approximation, $f_\trm{el}=f_\trm{PB}$. We solve the PB equation numerically in the plate--plate geometry. For charged plates immersed in solution of ions with concentration $c_i$ and valence $q_i$, the PB equation for the mean electrostatic potential $\psi(z)$ reads
\begin{equation}
\frac{{\rm d}^2 \psi(z)}{{\rm d} z^2} =  -\frac{e_0}{\varepsilon\varepsilon_0}\sum_i c_i \,\rme^{-q_ie_0\beta\psi(z)} ,
\label{eq:PB}
\end{equation}
where $z$ is the coordinate normal to the plates, $e_0$ is the elementary charge, $\varepsilon$ is the dielectric constant and $\beta = 1/(\kB T)$ is the inverse thermal energy. The PB equation is solved with the constant charge boundary conditions. Note that for symmetric systems the choice of boundary conditions does not have a big influence on the force~\cite{Trefalt2016}. The disjoining (osmotic) pressure $\Pi_{\rm PB}$ is then calculated from the first intergral of the PB equations at the mid-point between the bounding plates 
\begin{equation}
\Pi_{\rm PB} = \kB T\sum_i c_i (e^{-q_ie_0\beta\psi(0)}-1),
\label{eq:pressPB}
\end{equation}
where $\psi(0)$ is the midpoint potential at $z=0$. Integration of the osmotic pressure finally leads to the interaction free energy per surface area
\begin{equation}
f_{\rm PB} (h) = \int_h^{\infty} \Pi_\trm{PB}(h'){\rm d}h',
\label{eq:forcePB}
\end{equation}
which enters the total DLVO interaction free energy in Eq.~(\ref{eq:force}), namely
\begin{equation}
f_\trm{DLVO}(h) = f_{\rm PB}(h) + f_{\rm vdW}(h).
\label{eq:DLVO}
\end{equation}
General features of the PB interaction free energy are well known and have been investigate in minute details~\cite{Andelman}. 

\subsubsection{Dressed ion theory}
The theoretical approach that is the basis of this study, is the DI approximation. Technically, while being {\sl related} to the strong coupling approach~\cite{Moreira2001, Boroudjerdi}, the DI approximation differs from it in implementation details, since it is not devised to be an exact limiting fixed point, but rather corresponds to an {\sl approximate virial expansion} in terms of the small concentration of multivalent ions $c_0$, usually relevant in the experimental context~\cite{Kanduc2010, Kanduc2011,Perspective}. The formalism of the dressed ion approximation depends very crucially on the ensemble describing the multivalent counterions. While the strong-coupling approach is based on the canonical ensemble with fixed number of counterions, corresponding to the full neutralization of the system, the dressed counterion approximation is rather based on the grand canonical ensemble, since the multivalent salt is also in equilibrium with a bulk solution, leading to important quantitative changes in the results of both otherwise qualitatively related approaches. The grand canonical ensemble will also be the ensemble of choice when comparing the behavior of the DI theory with experiments below. In our implementation of DI theory, the polyvalent ions are confined between the charged surfaces, while the monovalent ions are assumed to be present in all the regions in space. This minor adjustment significantly simplifies the equations, yet it retains the qualitative picture of the linearized PB~\cite{Kanduc2010}.

{Formally, the DI interaction pressure is obtained from the grand canonical partition function for a Coulomb fluid with screened electrostatic interactions between multivalent ions, virialy expanded to the first order in the bulk fugacity of the multivalent salt in the presence of external fixed charges. While the latter assumption could be generalized, we do not venture into these further generalizations~\cite{Adzic1}. Because of the decoupling of the theory into the weakly coupled salt and dressed counterions, the interaction pressure similarly decomposes into a linearized PB (Debye--H\"{u}ckel) part, and a correlation part linear in the fugacity of the multivalent salt.}

The virial expansion of the electrostatic free energy per surface area in terms of the multivalent counterion concentration $c_0$ can be written as
\begin{equation}
f_\trm{el}=w_{00}+c_0 w_\trm{DI}^{(1)}+c_0^2 w_\trm{DI}^{(2)}+\ldots.
\label{eq:f}
\end{equation}
The first term, $w_\trm{00}$, is the `bare' surface--surface screened electrostatic repulsion on the linearized PB level,
\begin{equation}
w_{00}=\frac{\sigma^2}{2\varepsilon\varepsilon_0\kappa}\,\rme^{-\kappa h},
\end{equation}
where $\sigma$ is the surface charge density, $\kappa = [8 \pi \lB (n_0+\tfrac 12 q c_0)]^{1/2}$ is the inverse Debye screening length calculated from the concentrations of the monovalent ions from monovalent $n_0$ and multivalent $c_0$ salts, and $\lB = \beta e_0^2/(4\pi\varepsilon\varepsilon_0)$ is the Bjerrum length. The remaining terms in \Eq~(\ref{eq:f}) stem from contributions due to the multivalent counterions and can be traced to the corresponding terms in the virial expansion. 

In this work, we restrict ourselves to the contributions up to the second order in counterion concentration $c_0$, so that 
\begin{eqnarray}
\beta w_\trm{DI}^{(1)}&=&-\cZ_1'+h¸,\label{eq:wDI1}\label{eq:wDI1}\\
\beta w_\trm{DI}^{(2)}&=&\cB_2(h)-B_2(2\cZ_1'-h).\label{eq:wDI2}
\end{eqnarray}
The leading-order term $w_\trm{DI}^{(1)}$ is a one-particle, first order virial contribution, introduced already in Ref.~\cite{Kanduc2010}, related to the partition function $\cZ_1'$ of a single counterion between the surfaces,
\begin{equation}
\cZ_1'=\int_{-h/2}^{h/2}\rme^{-\beta u_\trm{0c}(z)}\,\rmd z.
\label{eq:Z1}
\end{equation}
The linearized PB interaction energy of the counterion with both charged surfaces $u_{0c}$ is given by~\cite{Kanduc2010}
\begin{equation}
\beta u_\trm{0c}(z) = -\frac{2}{\kappa\mu}\rme^{-\kappa h/2}\cosh \kappa z,
\end{equation}
where $\mu=e_0/2\pi q\lB \sigma$ is the Gouy--Chapman length. The one-particle level of description is valid as long as the multivalent salt concentration $c_0$ is small enough. As it becomes higher, also multivalent counterion--counterion interactions, $u_\trm{cc}$, become important.  On the linearized PB level it assumes the form
\begin{equation}
\beta u_\trm{cc}(r)=\frac{q^2 \lB}{r}\,\rme^{-\kappa r},
\end{equation}
where $r$ is the distance between the counterions. The two-particle, i.e. the second virial order, contribution is given by \Eq~(\ref{eq:wDI2}). Here, the dominating term is provided by the {\sl generalized second virial coefficient}
\begin{equation}
\cB_2(h)=-\frac 12\int f_{12}\,\rme^{-\beta u_\trm{0c}(z)-\beta u_\trm{0c}(z')}\rmd\boldsymbol{\rho}\rmd z\rmd z',
\label{eq:cB2}
\end{equation}
which is a weighted integral of the Mayer function $f_{12}=\exp(\beta u_\trm{cc}(\Av r))-1$ over the entire slit between the bounding surfaces, with $z$ and $z'$ being the surface-normal coordinates of the counterions and $\boldsymbol{\rho}$ the lateral vector between them. Furthermore,  $B_2$ is the standard (bulk) counterion--counterion second virial coefficient, defined as the volume integral over the Mayer function,
\begin{equation}
B_2=-\frac 12\int f_{12}\,\rmd \Av r.
\end{equation}
Detailed derivation of the second virial order contribution is given in Appendix~\ref{app:expansion}.

As it generally turns out, higher order virial terms quickly become unsuitable for practical implementation in the regime of strong many-body effects that can be fully implemented by taking recourse to computer simulations. We will therefore limit our calculations in this work only to the one-particle contribution $w_\trm{DI}^{(1)}$, whereas the second-order correction, $w_\trm{DI}^{(2)}$, will serve us as a qualitative insight into the many-body effects and to predict the regime of validity of the DI approach. 

The DI treatment is applicable when the second-order correction in \Eq~(\ref{eq:f}) is much smaller than the leading-order DI term, that is, when $\vert c_0 w_\trm{DI}^{(1)}\vert\gg \vert c_0^2 w_\trm{DI}^{(2)}\vert$. As derived in Appendix~\ref{app:criterion}, within an approximation of small surface separation $h$\,$\ll$\,$\kappa^{-1}$, the DI theory is applicable when
\begin{equation}
 c_0 \ll\frac{2\kappa^2}{\pi h} \frac{\exp\bigl(-\frac{2}{\kappa\mu}+\frac{h}{\mu} \bigr)}{\log^2 (q^2\lB\kappa)},
\label{eq:criterion}
\end{equation}
which depends on the surface separation $h$. This implies, that the one-particle DI theory works better for lower multivalent salt concentrations $c_0$, and for larger screening $\kappa$ (stemming predominantly from monovalent background salt). This can be explained as follows: due to the lack of repulsive ion--ion interactions in the one-particle DI approach, the theory overestimates the uptake of multivalent ions into the slit. Conversely, higher screening  reduces the attraction between multivalent ions and surfaces, resulting in a reduced amount of multivalent ions in the slit, consequently expanding the applicability of the one-particle approximation. The validity depends non-monotonically on the surface charge (reflected in the Gouy--Chapman length $\mu$).

The force $F$ between two spherical particles is then calculated through Derjaguin approximation ~\Eq~(\ref{eq:Derjaguin}), and can be expressed through the interaction free energy between planar surfaces.

\subsubsection{Dressed ion theory with effective potentials (surface charges) and additional attractions}

\begin{figure}[t]
\small
\centering
\includegraphics[width=8.5cm]{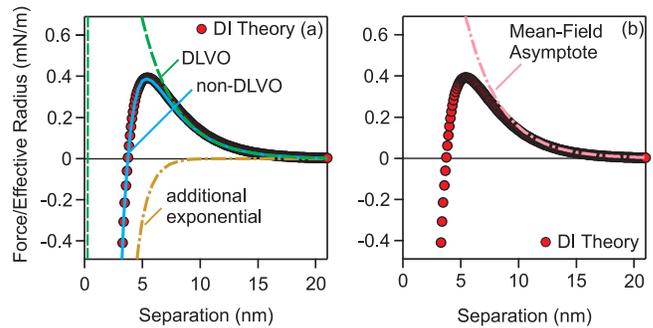}
\caption{Interpretation of the results from the DI theory where PB is used as a benchmark case. (a) Force {\it vs} separation between two charged surfaces where electrostatic part was calculated with the DI theory Eq.~(\ref{eq:f}). The DLVO and non-DLVO fits are also shown. (b) The same force shown together with the mean-field asymptote. The parameters $\sigma = 20$~mC/m$^2$, $n_0 = 10$~mM, $c_0 = 0.01$~mM, $q = 4$, $H = 3.5\times10^{-21}$~J were used for DI theory calculations.}
\label{fig:fittingDressed}
\end{figure}

In what follows, we now take the full PB theory as the benchmark case and try to capture the deviations from the mean-field approximation as derived from the DI theory. The force profiles calculated with the DI theory with vdW force (see above) included are shown in Fig.~\ref{fig:fittingDressed}a.
%%%%%%%%%%%%%%%%%%%%%%%%%%%%%
%%%%%%%%%%%%%%%%%%%%%%%%%%%%%
At large separations, the DI curve can be accurately fitted with the mean-field PB plus vdW (DLVO) theory. The fitted surface potential $\psi_{\rm eff} = 37$~mV, however, does not correspond to the actual surface charge but is an {\sl effective parameter}. Conversely, at small separations the DI theory predicts a strong attraction, which is not captured by the mean-field approximation. In order to quantify the deviation of the DI theory from the PB baseline in this regime of separations, we introduce a modified non-DLVO phenomenological form of the interaction to describe this additional attraction
\begin{equation}
f_{\rm nonDLVO} = f_{\rm DLVO} - A\,\rme^{-h/\lambda} ,
\label{eq:nonDLVO}
\end{equation}
where $\lambda$ and $A$ are the decay length and the amplitude of this additional interaction. This choice of the phenomenological non-DLVO interaction form, Eq.~(\ref{eq:nonDLVO}), is obvisouly capable to describe the experimental force profile rather accurately. The fit of this additional exponential force shown in Fig.~\ref{fig:fittingDressed}a yields $A = 45$~mN/m and $\lambda = 1$~nm.

The interpretation of these observations is the following: at large separations the electrostatic potential is rather small and the mean-field behavior is sufficient to characterize the experiment but only if in addition one takes into account that electrostatic interactions between the surface and the counterions reduce the effective surface charge from its bare PB value. At small separations the situation is altogether different, and ion correlations induce strong attraction, which cannot be captured by the mean-field {\sl Ansatz} but can be approximated by the phenomenological attractive exponential form that, as we will show, can be rationalized by the DI approach.

The long-distance mean-field behavior equivalently results directly from the asymptotic limit of the DI theory. In the limit of large separations, $\kappa h \gg 1$, Eq.~(\ref{eq:f}) reduces to 
\begin{equation}
f_{\rm el}(h) \simeq  w_{00}(h) K(h),
\label{eq:DCasym}
\end{equation}
which exhibits exponentially decaying interaction, mildly modulated by the function $K(h)$ of the form
\begin{equation}
K (h) =  1- c_0 \frac{2\pi\lB q^2}{\kappa^2}\Bigl( C+\kappa h\Bigr),
\label{eq:DCasymConst}
\end{equation}
with the constant $C$ given by
\begin{equation}
C= 3/2 -2\gamma+  2\log{\kappa\mu}+2\kappa\mu\,\rme^{1/\kappa\mu} +2\trm{Ei}(1/\kappa\mu).
\label{ncbrjsfg}
\end{equation}
Here,  $\gamma$\,$=$\,$0.577\ldots$ is the Euler's constant and $\trm{Ei}(x)$ the exponential integral. The derivation is given in Appendix~\ref{app:asymptotic}. In our experimental cases, the product $\kappa\mu$ is typically below 0.2, and therefore for $\kappa\mu$\,$\ll$\,1, we can make the approximation $C\simeq 2(\mu\kappa)^2 \exp(1/\kappa\mu)$. The value of $K$ then determines the effective potential, which assumes the form
\begin{equation}
\psi_{\rm eff} = \frac{\sigma}{2\varepsilon\varepsilon_0 \kappa}\sqrt{K}.
\label{eq:effPot}
\end{equation}
The asymptote corresponding to this result is plotted in Fig.~\ref{fig:fittingDressed}b and can be seen to nicely describe the long-distance exponential behavior, which suggests that the long-distance mean-field fitting is reasonable and accurate. At the experimental conditions $\kappa \mu \sim 0.1$, so that $\kappa h$ is negligible, and only the last two terms in Eq.~(\ref{ncbrjsfg}) are important, while the rest can be neglected.

\subsection{Experimental}

Experiments were carried out with two sets of colloidal particles, mono-dispersed silica particles (Bangs Labs) and polystyrene sulfate latex~(SL) particles (Invitrogen). Note that both types of particles are negatively charged in pH~4 water, therefore multivalent cations were used as counterions.

The silica particles were attached to the cantilever and a substrate by heat treatment at 1150~\celsius\ for 3~h. After the heat treatment the particles were characterized. The average particle radius of 2.20~\si\um\ and a coefficient of variation (CV) of 1.2~\% was measured with scanning electron microscopy. The root mean square (rms) roughness of $0.81\pm0.09$~nm was obtained by AFM imaging. The forces between two silica spheres were then measured by colloidal probe technique with a closed-loop AFM (MFP-3D, Oxford Instruments). The particles were immersed in LaCl$_3$, pH 4.0 solutions within the AFM liquid cell, which was placed on the inverted optical microscope. The optical microscope enables the centering of the two spheres with the precision of $\sim$50~nm. The forces were obtained from the approach-retract cycles with cantilever velocities of 0.3~\si\um/s. The average force profiles were obtained by averaging about 150 approach cycles while the deflection was converted to force with the Hook's law using the measured value of the cantilever spring constant. Further details on the measurements with silica particles are given in Ref.~\cite{Valmacco2016}.

The forces between two negatively charged sulfate latex spheres of 3~\si\um\ (CV 4.1~\%) were measured in aqueous solutions of two different oligoamines. In particular, we used chloride salt of triethylenetetramine H$_2$N(CH$_2$CH$_2$NH)$_2$CH$_2$CH$_2$NH$_2$ (N4), and the basic form of pentaethylenehexamine H$_2$N(CH$_2$CH$_2$NH)$_4$CH$_2$CH$_2$NH$_2$ (N6) both purchased from Aldrich. The pH was adjusted to 4.0 with HCl and KOH. The background salt level was adjusted with KCl. At these conditions the average valence of the N4 and N6 species in the solution is $+3$ and $+4$, respectively. The force measurements with the sulfate latex particles were done in similar fashion as with the silica particles. The main difference in the method was the attachment of the particles to the cantilever and the substrate. 

While in the case of silica particles they were attached with the heat treatment, the latex particles were attached {\sl in situ} in the fluid cell. For the attachment the cantilever and the substrate was silanized with hexamethyldisilazane (Alfa Aesar). The latex particles were then left to deposit on the glass substrate at the bottom of the fluid cell and picked up by the tip-less cantilever by pressing the particle against the substrate. After alignment of the two particles the approach--retract cycles were recorded. The deflection of the cantilever was converted to force using the same procedure as for the silica particles. Further details on the measurements with sulfate latex particles can be found in Ref.~\cite{MoazzamiGudarzi2015}.

Note that majority of experimental results shown in this manuscript were already presented in Refs.~\cite{Valmacco2016, MoazzamiGudarzi2015}, while some new experimental data involving higher background monovalent salt are presented here.

\section{Results and Discussion}

In this study, symmetric silica--silica and SL--SL interactions are measured with colloidal probe AFM in the presence of multivalent counterions. Both types of particles are negatively charged in water at pH~4.0, immersed in an electrolyte mixture with monovalent salt and multivalent cations. In the case of silica, interaction forces in the presence of La$^{3+}$ ions were studied. For SL particles, oligoamines N4 and N6 with the average valencies of $+3$ and $+4$ were used. The generic features of such interaction forces show {\sl repulsion} at low multivalent salt concentration with decreasing magnitude upon increasing the multivalent salt concentration. The effective potential shows a similar trend until the charge neutralization point is reached, where the forces then turn {\sl attractive} due to the underlying vdW~(\ref{eq:vdW}) and non-DLVO~(\ref{eq:nonDLVO}) attractions. Upon further addition of the multivalent ions, in some situations the overcharging or charge-reversal occurs, which again results in repulsive forces~\cite{MoazzamiGudarzi2015}. All these forces exhibit mean-field behavior at large distances and additional, non-mean field, attractions at small distances.

We now focus on comparison between experimental force curves and predictions of the DI theory. The DI force curve (Fig.~\ref{fig:fittingDressed}) can be decomposed into a {\sl long-range component}, where the mean-field behavior with an effective potential is recovered, and a {\sl short-range component}, where deviation from the mean-field predictions can be described by an additional attractive term of an exponential type, see \Eq~(\ref{eq:nonDLVO}). Similar decomposition can be devised also for the experimental curves, and has been used earlier~\cite{MoazzamiGudarzi2015, MontesRuiz-Cabello2013, Valmacco2016, Moazzami-Gudarzi2016}.

\begin{figure}[t]
\small
\centering
\includegraphics[width=8.5cm]{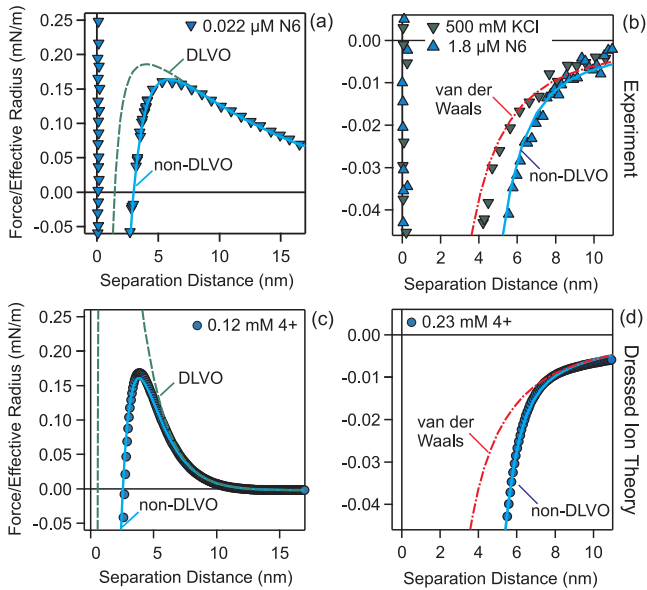}
\caption{(a), (b) Experimentally measured forces between two SL particles in the presence of tetravalent N6. (c), (d) Forces calculated with the DI theory in the presence of tetravalent counterions. For both cases forces at lower counterion concentration (left) and at charge neutralization (right) are shown. The DLVO Eq.~(\ref{eq:DLVO}) and non-DLVO Eq.~(\ref{eq:nonDLVO}) fits are plotted as well. The parameters $\sigma = 24$~mC/m$^2$, $n_0 = 25$~mM, $c_0 = 0.12$~mM and $c_0 = 0.23$~mM, $q=4$, and $H = 3.5\times10^{-21}$~J, the latter determined at 1 M concentration of KCl.}
\label{fig:expDressed1}
\end{figure}

In Fig.~\ref{fig:expDressed1}a the measured and calculated forces for tetravalent counterions are juxtaposed, 
showing forces between two SL particles in the presence of N6 with 1~mM background 1:1 electrolyte. At low concentration (0.022~\si\micro M) of N6, the force is repulsive at large distances. The repulsion is greatly reduced as compared to the case without N6, with only 1 mM of the background salt (not shown here) and the magnitude of the effective potential is reduced from $\sim$ 80~mV to $\sim$20~mV~\cite{MoazzamiGudarzi2015}. At small intersurface distances the experimental curve is more attractive than predicted by the DLVO theory. This augmented attraction can be well fitted with the negative exponential term in Eq.~(\ref{eq:nonDLVO}) with a decay length of $\lambda= 1.0$~nm and an amplitude of $A = 1.5$~mN/m. For the present calculations the Hamaker constant of $H = 3.5\times10^{-21}$~J was used, which was determined at high concentration of monovalent salt, where the electrostatic force is completely screened and only the vdW force is present~\cite{MoazzamiGudarzi2015}.

In Fig.~\ref{fig:expDressed1}b the force curve at charge neutralization is shown, which occurs at 1.8~\si\micro M of N6. The force is attractive in the whole distance range, however this attraction is stronger than the vdW attraction predicted by the DLVO theory. The vdW force is recovered at high concentrations of monovalent KCl salt and this profile is also shown for comparison. The additional attraction induced by N6 can be again described by the exponential term with the same decay length of $\lambda=1.0$~nm but larger amplitude of $A = 4.1$~mN/m. 

In Figs.~\ref{fig:expDressed1}c,~d forces calculated with the DI theory at similar conditions are shown. At low concentration of tetravalent counterions (left) the force is repulsive at large distances and is consistent  with the mean-field description. At short distances an additional attraction emerges again, which can be fitted using Eq.~(\ref{eq:nonDLVO}) with $\lambda =0.7$~nm and $A = 28$~mN/m. When the concentration of multivalent counterions is increased (Fig.~\ref{fig:expDressed1}d), the force turns attractive. This attraction is stronger than the bare vdW force, because it additionally stems also from the electrostatic part, $f_\trm{el}$. 

When the experimental and the DI force curves are compared, the same qualitative behavior is revealed in both cases; compare top and bottom panels of Fig.~\ref{fig:expDressed1}. Although the comparison is not fully quantitative, since the $c_0$ and $n_0$ concentrations were adjusted for the calculation, viz.\ the $n_0$ was increased in order for $c_0$ to fall below the DI validity bound, see~\Eq~(\ref{eq:criterion}), which in this particular case amounts to 3 mM at distance of 3 nm. Nevertheless, with the adjusted parameters, the DI theory correctly predicts the decrease in the magnitude of the effective potential and, even more importantly, it {\sl captures the range and the magnitude of the additional attraction}.  The good agreement suggests that the additional attraction observed in the experimental curves is caused by the ion-correlation effects as they enter the DI formalism. However, if one uses directly the experimental concentrations and inserts them into the DI theory, this yields an overestimation of the additional attraction. The disagreement comes from the fact that DI theory neglects the repulsive interactions between multivalent ions and thus overestimates their density in the slit, leading to too large attraction between the surfaces, which has been demonstrated also by comparing DI results with Monte Carlo simulations~\cite{Kanduc2010}.
 
\begin{figure}[h!]
\small
\centering
\includegraphics[width=8.5cm]{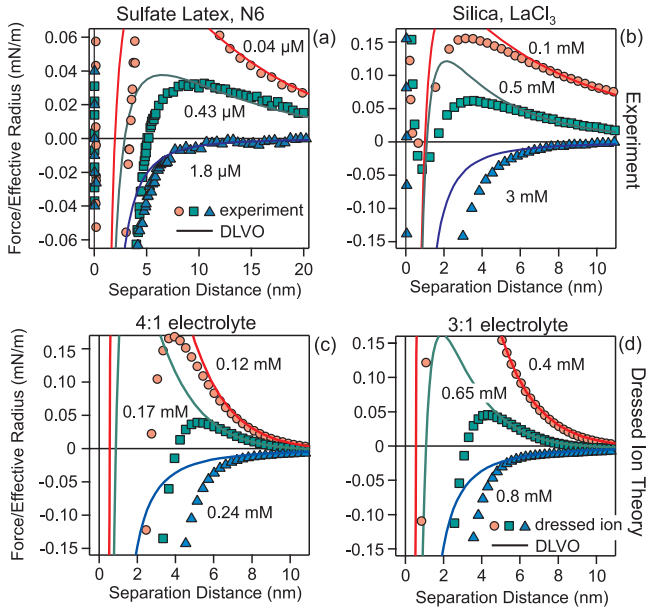}
\caption{Force evolution with increasing multivalent counterion concentration. (a) Forces between SL particles in the presence of tetravalent N6 ions. (b)~Forces between silica particles in the presence of trivalent La$^{3+}$ counterions. (c)~DI theory force profiles for tetravalent counterions. (d)~DI theory force profiles for trivalent counterions. DLVO fits are shown for all profiles as solid curves. Parameters used for DI theory are $\sigma = 24$~mC/m$^2$, $n_0 = 25$~mM, $H = 3.5\times10^{-21}$~J, $q = 4$ for (c) and $q = 3$ for (d), and $c_0$ as noted in the figures.
}
\label{fig:expDressed2}
\end{figure}

Let us now focus upon the evolution of the forces with increasing concentration of the multivalent counterions in more detail. In Fig.~\ref{fig:expDressed2} we again present a comparison between experimental (a), (b) and calculated (c), (d) force curves.
The interactions between two SL particles in the presence of tetravalent counterions are shown in Fig.~\ref{fig:expDressed2}a, while similar interactions for silica particles in the presence of trivalent counterions are shown in Fig.~\ref{fig:expDressed2}b. As in Fig.~\ref{fig:fittingDressed} all the force curves can be decomposed into a long-range part approaching the mean-field predictions,  and a short-range part exhibiting an additional attraction. With increasing counterion concentration the forces become less repulsive and finally turn totally attractive at the charge neutralization point. Accordingly, the magnitude of the effective potential decreases with increasing concentration. In all the measured force curves, an additional attraction at distances below $\sim$\,5~nm is clearly present. The bottom row of Fig.~\ref{fig:expDressed2} presents the DI calculations for similar conditions, namely tetravalent ions left and trivalent ions right. Again one can see that the DI theory force curves closely resemble the experimental curves, with both the long-distance mean-field behavior as well as the additional short range attraction. Moreover, the DI theory also correctly predicts the decrease of the magnitude of the effective potential (see below) with increasing multivalent counterion concentration, as well as, importantly, that lower concentrations of tetravalent as compared to trivalent counterions are needed to reach charge neutralization.

\begin{figure}[ht]
\small
\centering
\includegraphics[width=8.5cm]{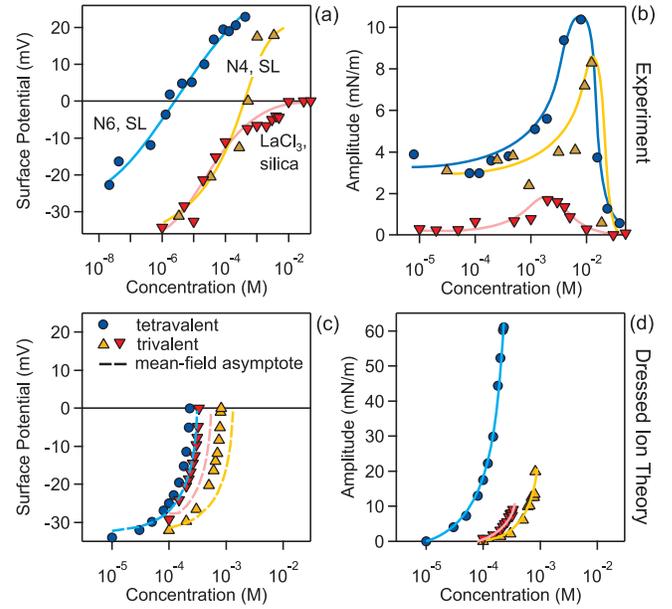}
\caption{Effective surface potentials (left) and the strength of the additional attractive force (right) {\it vs.} the concentration of multivalent counterions. (a), (b) Experimental values for SL particles in the presence of N4 and N6, and silica particles in the presence of La$^{3+}$ ions. (c), (d) DI theory results for higher surface charge $\sigma = 24$~mC/m$^2$, $n_0 = 25$~mM with trivalent and tetravalent counterions and smaller surface charge $\sigma = 15$~mC/m$^2$, $n_0 = 10$~mM with trivalent counterions. For effective potentials in (c) asymptotic limits Eq.~(\ref{eq:effPot}) are also shown by dashed lines.}
\label{fig:expDressedParameters}
\end{figure}

The forces presented in Fig.~\ref{fig:expDressed2} were further analyzed by fitting the Eq.~(\ref{eq:nonDLVO}). This analysis yields three parameters: the effective potential, $\psi_{\rm eff}$, the amplitude, $A$, and range, $\lambda$ of the phenomenological exponential attraction. Note that $\psi_{\rm eff}$ is obtained by the numerical solution of the PB equation, Eq.~(\ref{eq:PB}), where it enters as a boundary condition, for details see Ref.~\cite{MontesRuiz-Cabello2013}. The range of the additional attraction $\lambda$ is practically independent of the concentration and is therefore fixed to its average value. The remaining two parameters, $\psi_{\rm eff}$  and $A$, are summarized in Fig.~\ref{fig:expDressedParameters}.
Again the experimental data are shown in the top panel, Fig.~\ref{fig:expDressedParameters}a,~b, and DI theory results in the bottom panel, Fig.~\ref{fig:expDressedParameters}c,~d. Effective potentials extracted from the measured interactions between SL particles in the presence of trivalent and tetravalent counterions and silica particles in the presence of trivalent ions show similar behavior. The potentials are negative at low counterion concentrations and increase with addition of multivalent counterions until the charge neutralization point. For the case of SL particles and oligoamines, the potentials then flip to positive values and overcharging is observed, which causes again a repulsive interaction in the force profiles as shown in Ref.~\cite{MoazzamiGudarzi2015}. Charge neutralization point is shifted to lower concentrations when the valence of the oligoamine is increased. Fig.~\ref{fig:expDressedParameters}b shows the evolution of the strength of the additional attraction with the counterion concentration. For all three experimental cases the additional attraction is small at low concentration and increases with increasing concentration, reaching a maximum, and subsequently decreases again at high concentrations. The strength of the additional attraction is more pronounced and the maximum is shifted to lower concentrations for the higher valence oligoamine N6. 

Similar trends as the ones above are observed also within the DI theory, see Fig.~\ref{fig:expDressedParameters}c,~d. The predicted effective potentials are negative for low concentrations and increase with increasing concentration until the charge neutralization is reached, a behavior completely consistent with the experimental observations. The shift of the neutralization point to lower concentrations with higher valency is also correctly predicted. Moreover, the effective potentials calculated with the asymptotic limit Eq.~(\ref{eq:effPot}) agree very well with the fitted DLVO theory. The strength of the additional attraction predicted by the DI theory is shown in Fig.~\ref{fig:expDressedParameters}d. The decay lengths of the additional attraction of 0.7~nm and 1.1~nm for the higher and the lower surface charge, respectively, closely match the experimentally determined values of $\lambda = 1.0$~nm. The amplitude of the additional force increases with increasing concentration in a similar fashion to experimentally determined values. A higher amplitude as well as a shift to lower concentrations for higher valency are also correctly captured by the DI theory. 

On the other hand, the DI theory is unable to predict the effective potentials beyond charge neutralization. The experimental forces show {\sl re-entrant behavior}, they are repulsive at low concentrations, attractive at charge neutralization, and become repulsive again at higher concentrations due to overcharging, for details see Ref.~\cite{MoazzamiGudarzi2015}. The DI theory correctly captures the first part: the calculated forces are repulsive at low concentrations and get attractive at charge neutralization. However, upon further addition of multivalent ions the forces stay attractive and no repulsion is recovered. This latter mismatch between the experiment and the DI theory is due to the fact that the repulsive interactions between the multivalent ions are not implemented in the DI theory, and they become important at these conditions. The maximum and the turn-over in the trend of the strength of the additional attractive force is not captured for the same reason. These problems aside, the DI theory correctly predicts the overall behavior below the charge neutralization point. 

Different implementations of the strong coupling phenomenology of multivalent cations, emphasizing the strong Wigner crystal-like correlations of multivalent cations at the charged surface, which have been implemented to describe the re-entrant DNA condensation \cite{Nguyen2} could possibly be appropriate to shed light on this re-entrant behavior of the colloidal interactions.

\begin{figure}[t]
\small
\centering
\includegraphics[width=6.5cm]{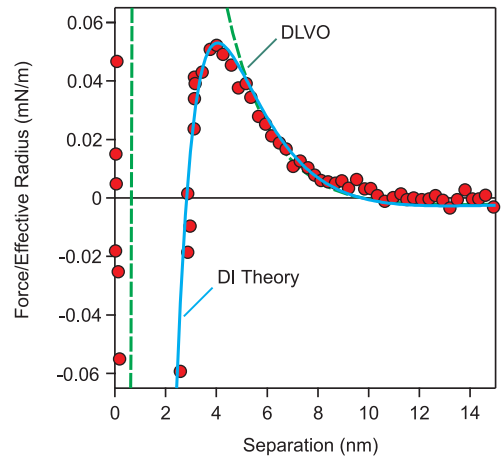}
\caption{Direct comparison of experiments and DI theory. The interaction between two SL particles is measured in the presence of 4.3 \si\micro M N6 and 23 mM KCl background electrolyte. Predictions from the DI theory and DLVO theory are also shown. Parameters for the DI theory obtained from the fits are $\sigma = 18$~mC/m$^2$, and $c_0 = 0.265$~mM, while $n_0 = 23$~mM,  $q=4$, and $H = 3.5\times10^{-21}$~J are fixed to experimental values.}
\label{fig:expDressedHighBackground}
\end{figure}

At higher concentrations of 1:1 background salt, the DI theory can be directly compared with the experimental force profile, see Fig.~\ref{fig:expDressedHighBackground}. 
Here the force between two SL particles in the presence of N6 is measured, with the concentration of the background 1:1 electrolyte increased from 1 mM to 23 mM. The DLVO fit yields an effective potential at larger separation, but similarly to previous cases, the mean-field theory is not able to explain the additional attraction at smaller distances even with allowance for effective parameters to be fitted from experiments. On the other hand, the DI theory works quantitatively and with parameters obtained from experiments captures the main features of the force profile. Furthermore, the experimental values for the concentration of the background electrolyte, valency and the Hamaker constant are used for the DI theory prediction~\cite{MoazzamiGudarzi2015}. The only adjustable parameters are surface charge density and the concentration of the multivalent counterion. The fitted surface charge density of 18~mC/m$^2$ matches very well with the experimental value of $24\pm7$~mC/m$^2$ determined from the measurements in KCl solutions~\cite{MoazzamiGudarzi2015}. The validity criterion \Eq~(\ref{eq:criterion}) for these set of parameters and the separation of $h=2.5$~nm, where the attraction occurs, implies the DI prediction to be valid for $c_0\ll$\,1~mM, which is a condition fulfilled in the experiments.
On the other hand, the multivalent counterion concentration is predicted to be significantly higher than the nominal concentration used in the experiment (Exp: 0.0043~mM, DI: 0.265~mM). The overestimation of the multivalent counterion concentration could be caused by non-electrostatic specific interactions of the N6 with the surface, leading to charge regulation of the surface charge, driven by the multivalent counterions. Such non-electrostatic interactions not accounted for in the DI theory could enhance the adsorption and consequently decrease the bulk concentration of multivalent ions needed to reach the same effective charge. On the other hand, charge regulation of the multivalent counterions themselves, or bridging of the counterion chains between the surfaces could also play a role here. Our experiments simply do not allow to clearly pinpoint the exact mechanism at this point.

\section{Conclusions}

In this work, we have compared the predictions of the dressed ion theory with experimentally measured force profiles between negatively charged particles in the presence of multivalent cations. Such forces exhibit two distinct behaviors at large and small separations, respectively. At large separations, they can be well fitted by the mean-field PB theory, yielding effective values for the surface potentials. This result is confirmed by the DI theory, which in the asymptotic limit of large separations also captures the mean-field behavior and is in addition capable of predicting the effective potentials. At shorter separations, the mean-field predictions qualitatively deviate from the experimental forces and an additional attraction is evident. This attraction can be approximated by an additional phenomenological exponential term, which then phenomenologically corrects the mean-field DLVO approach. 

On the other hand, the DI theory successfully predicts the quantitative deviation from the mean-field result and predicts the values of the phenomenological parameters of the additional short-range attraction. The predicted range of the decay length $\lambda\sim 1$~nm as well as the strength of the additional attraction are both very well within the experimental values. This suggests that the additional attraction is probably caused by ion correlations and that the level of formalization of these effects implied by the DI theory is adequate. Furthermore, the DI theory also successfully describes the dependencies of the effective potential (surface charge) and the strength of the additional attraction on the counterion concentration and valency. Although at lower concentrations of background 1:1 salt ($<1$~mM) the predictions are not quantitative, qualitative behavior up to the charge neutralization is captured perfectly. 
At higher concentrations of background salt ($\sim 20$~mM), the DI theory describes the force profile quantitatively with only one adjustable parameter, {\it i.e.} the effective concentration of the multivalent counterions. 
The better predictions at higher background salt concentrations are due to reduced repulsive counterion--counterion interactions, as corroborated by the second-order virial expansion of the DI approach.
The remaining adjustment of the counterion concentration is possibly due to non-electrostatic ion-specific effects~\cite{Yaakov}, which have not been explicitly taken into account.

\section*{Acknowledgments}
GT, MMG and VV acknowledge the Swiss National Science Foundation and the University of Geneva for the financial support. RP acknowledges the financial support of the Agency for research and development of Slovenia (ARRS) through grants J1-7435 and P1-0055.

\appendix
\section{Virial expansion}
\label{app:expansion}
Here we derive the free energy contribution of DI to the total free energy of the system to the second order in the virial expansion given by \Eq~(\ref{eq:f}).

The exact form of the grand canonical free energy $\cF_\trm{DI}^0$ of DI gas between the surfaces can be expressed in terms of multivalent ions fugacity $\Lambda_\trm{c}$ as
\begin{equation}
\rme^{-\beta \cF_\trm{DI}^0}=\sum_{N=0}^\infty \Lambda_\trm{c}^N \cZ_N,
\end{equation}
where $\cZ_N$ are the canonical partition functions of $N$ ions, with the lowest two explicitly expressing as
\begin{eqnarray}
\cZ_1&=&\frac{1}{1!}\int \rme^{-\beta u_\trm{0c}(\Av r)}\rmd \Av r,\\
\cZ_2&=&\frac{1}{2!}\int \rme^{-\beta (u_\trm{0c}(\Av r)+u_\trm{0c}(\Av r')+u_\trm{cc}(\Av r-\Av r'))}\rmd \Av r\rmd \Av r'.
\end{eqnarray}
Here, $u_\trm{0c}(\Av r)$ and $u_\trm{cc}(\Av r-\Av r')$ are the surface--ion and ion--ion interactions, respectively, defined in the main text. 
The free energy expanded up to the second order in $\Lambda_\trm{c}$ equals to
\begin{equation}
\beta \cF_\trm{DI}^0=-\Lambda_\trm c \cZ_1 -\tfrac 12 \Lambda_\trm c^2(2\cZ_2-\cZ_1^2).
\label{eq:FDI0}
\end{equation}
The free energy of the bulk reservoir $\cF_\trm{DI}^{\trm{(bulk)}}$ can be obtained from  the above expression by setting $u_\trm{0c}=0$ and integrating throughout the entire space, viz.
\begin{equation}
\beta \cF_\trm{DI}^{\trm{(bulk)}}=-\Lambda_\trm{c}V +\Lambda_\trm{c}^2 B_2 V.
\label{eqA:FDI}
\end{equation}
This enables us to evaluate the concentration of multivalent ions $c_0$ in the bulk
\begin{equation}
c_0=-\frac{\Lambda_\trm c}V \left(\frac{\partial\beta \cF_\trm{DI}^{\trm{(bulk)}}}{\partial {\Lambda_\trm c}}\right)_{\beta,V},
\end{equation}
which is related to the fugacity $\Lambda_\trm{c}$ up to the second order as
\begin{equation}
{\Lambda_\trm c}=c_0+2B_2 c_0^2.
\label{eqA:muc0}
\end{equation}
The free energy of the bulk reservoir thus expresses with the concentration as
\begin{equation}
\beta \cF_\trm{DI}^{\trm{(bulk)}}=-c_0 V-c_0^2 B_2 V.
\label{eqA:Fbulk}
\end{equation}
The DI free energy of the entire system is a sum of the slit (\ref{eq:FDI0}) and the bulk (\ref{eqA:Fbulk}) contributions,  $\cF_\trm{DI}=\cF_\trm{DI}^0+\cF_\trm{DI}^\trm{(bulk)}$. The volume of the bulk reservoir $V$ is related to the slit volume as $V=V_0-Ah$, where $V_0$ represents the total volume of the system, which is a constant and can be omitted from further equations. The total DI free energy is then
\begin{eqnarray}
\beta \cF_\trm{DI}&=&-c_0(\cZ_1-Ah)\label{eqA:cFDI}\\
	&&+c_0^2 \left[\tfrac 12 \cZ_1^2-\cZ_2 -B_2 (2\cZ_1-Ah)\right],\nonumber
\end{eqnarray}
which expresses per surface area $f_\trm{DI}=\cF_\trm{DI}/A$ as
\begin{equation}
\beta f_\trm{DI}=-c_0(\cZ_1'-h)+c_0^2 \left[\cB_2(h) -B_2 (2\cZ_1'- h)\right].
\label{eqA:cfDI}
\end{equation}
In the last step, we have introduced surface-rescaled partition functions $\cZ_i'=\cZ_i/A$ and the generalized second virial coefficient,
\begin{equation}
\cB_2(h)=\left(\tfrac 12 \cZ_1^2 -\cZ_2\right)/A,
\end{equation}
given by \Eq~(\ref{eq:cB2}).
Equation~(\ref{eqA:cfDI}) represents the DI part of \Eq~(\ref{eq:f}).

\section{DI validity criterion}
\label{app:criterion}
In order to establish a simple criterion that can assess the regime in which the one-particle DI theory is valid, we need to compare the size of the one-particle contribution in \Eq~(\ref{eq:f}) to the next order. Thus, the one-particle DI theory is valid when $\vert c_0 w_\trm{DI}^{(1)}\vert\gg \vert c_0^2 w_\trm{DI}^{(2)}\vert$, which, after keeping the largest terms, approximately corresponds to
\begin{equation}
\cZ_1'\gg c_0 \cB_2(h).
\label{eq:crit1}
\end{equation}
In the following, we will focus only on small surface--surface separations, $h\ll \kappa^{-1}$, such that 
\begin{equation}
\cZ_1'\simeq h\,\rme^{-\beta u_\trm{0c}(0)}\simeq h\exp\left(\frac{2}{\kappa\mu}-\frac{h}{\mu}\right).
\end{equation}
Furthermore, in this limit the normal components $z$ and $z'$ in \Eq~(\ref{eq:cB2}) can be decoupled from the lateral component $\boldsymbol{\rho}$,
\begin{equation}
\cB_2(h)\simeq-\frac{1}{2}\cZ_1'^2\int_0^\infty f_{12}\,2\pi\rho\rmd\rho.
\end{equation}
In order to analytically evaluate the above integral, we approximate the Mayer function by a step function, being $f_{12}=-1$ for $\beta u_\trm{cc}>1$ and $0$ for $\beta u_\trm{cc}<1$.
The distance $\rho_0$ at the step, $\beta u_\trm{cc}(\rho_0)=1$, fulfills the condition
\begin{equation}
\kappa \rho_0\, \rme^{\kappa \rho_0}=q^2\lB \kappa,
\label{eq:rho0}
\end{equation}
and cannot be evaluated in a closed mathematical form. Since typically $q=3$ or $4$ and $\kappa=0.2$--$2$\,nm$^{-1}$, the factor in the right hand side of the above equation spans $q^2\lB \kappa=1$--$20$, therefore according to \Eq~(\ref{eq:rho0}) the factor $\kappa \rho_0$ is typically larger than unity. Thus, the exponent in the left-hand-side represents the dominating contribution, such that we can approximate $\kappa\rho_0=\log (q^2\lB \kappa)-\log{(\kappa\rho_0)}\simeq \log (q^2\lB \kappa)$.
This yields
\begin{equation}
\cB_2(h)\simeq\frac{\pi}{2\kappa^2} \cZ_1'^2\,\log^2 (q^2\lB\kappa).
\end{equation}
The validity criterion stemming from \Eq~(\ref{eq:crit1}) then yields
\begin{equation}
 c_0 \ll\frac{2\kappa^2}{\pi h} \frac{\exp\bigl(-\frac{2}{\kappa\mu}+\frac{h}{\mu} \bigr)}{\log^2 (q^2\lB\kappa)},
\end{equation}
which depends on the surface separation $h$ (valid only for $h\ll \kappa^{-1}$). 
Considering the most strict condition applicable to all separations, where the right-hand-side of the above expression reaches its minimum, which occurs for $h=\mu$, and after omitting numerical prefactors, we arrive at the criterion
\begin{equation}
c_0 \ll\frac{\kappa^2}{\mu} \frac{\rme^{-\frac{2}{\kappa\mu}}}{\log^2 (q^2\lB\kappa)}.
\end{equation}

\section{Asymptotic expression of $w_\trm{DI}^{(1)}(h)$}
\label{app:asymptotic}
We are interested in the long-range asymptotic behavior of the first-order DI interaction given by \Eq~(\ref{eq:wDI1}).
In the limit $\kappa h$\,$\gg$\,1, we can expand the partition function in \Eq~(\ref{eq:Z1}) as
\begin{equation}
\cZ_1'=2\int_0^{h/2}\exp\left(\frac{1}{\kappa\mu}\rme^{-\kappa(h/2-z)}\right)
\left[1+\frac{1}{\kappa\mu}\rme^{-\kappa(h/2+z)}\right]\rmd z .
\end{equation}
In order to evaluate the above integral, which involves mathematical functions of the form  $\exp(\rme^x)$, we use 
the exponential integral, Ei$(x)$, defined as 
\begin{equation}
\trm{Ei}(x)=-\int_{-x}^{\infty}\frac{\rme^{-t}}{t}\,\rmd t.
\end{equation}
Substituting $t=-\rme^{x}$, we arrive at 
\begin{equation}
\int\exp(\rme^x)\rmd x=\trm{Ei}(\rme^x).
\end{equation}
After the integration, we keep only the terms up to the order $\rme^{-\kappa h}$. To that end,
we use the Taylor expansion of $\trm{Ei}(x)$ for $x\ll1$, which is
\begin{equation}
\trm{Ei}(x)\simeq\gamma+\log x+x+\frac 14 x^2.
\end{equation}
We thus obtain the asymptotic expression for $\kappa h$\,$\gg$\,1 as
\begin{eqnarray}
\beta w_\trm{DI}^{(1)}(h)=-\frac{\rme^{-\kappa h}}{\kappa^3\mu^2}\Bigl(&&\kappa h +3/2-2\gamma+2\kappa\mu\,\rme^{1/\kappa\mu}\nonumber\\
&&+\>2\trm{Ei}(1/\kappa\mu)+2\log{\kappa\mu}\Bigl).
\end{eqnarray}
Using the derived expression in the one-particle free energy, gives as the expression \Eq~(\ref{eq:DCasym}).

\bibliography{paperLib}

\bibliographystyle{apsrev4-1}

\end{document}